\documentclass{article}
\usepackage{waspaa21,amsmath,graphicx,url,times}
\usepackage{color}
\usepackage{hyperref}  %
\usepackage{microtype}
\usepackage{amsfonts}       %
\usepackage{booktabs}       %
\usepackage{bm}
\usepackage{tabularx}
\usepackage{multirow}

\usepackage{enumitem}
\usepackage{nicefrac}

\setlist{nolistsep}

\newif\ifincludeappendix
\includeappendixfalse

\newif\ifincludeacknowledgements
\includeacknowledgementsfalse

\newif\ifarxiv
\arxivtrue

\ifarxiv
\includeacknowledgementstrue
\fi

\newcommand{\defeq}{\stackrel{\scriptscriptstyle\mathrm{def}}{=}}
\newcommand{\op}[1]{\operatorname{#1}}
\newcommand{\maxsnr}{\mathrm{SNR}_\mathrm{max}}

\title{Sparse, Efficient, and Semantic Mixture Invariant Training:\\
Taming In-the-Wild Unsupervised Sound Separation}

\name{Scott Wisdom, Aren Jansen, Ron J. Weiss,  Hakan Erdogan, John R. Hershey}
\address{Google Research\\{\small \texttt{\{scottwisdom,arenjansen,ronw,hakanerdogan,johnhershey\}@google.com}}}

\begin{document}

\ninept
\maketitle

\begin{sloppy}

\begin{abstract}
Supervised neural network training has led to significant progress on single-channel sound separation. 
This approach relies on ground truth isolated sources, which precludes scaling to widely available mixture data and limits progress on open-domain tasks.
The recent \emph{mixture invariant training} (MixIT) method enables training on in-the-wild data; however, it suffers from two outstanding problems.  First, it produces models  which tend to over-separate,  producing more output sources than are present in the input.  Second, the exponential computational complexity of the MixIT loss limits the number of feasible output sources.  
In this paper we address both issues. To combat over-separation we introduce new losses: \emph{sparsity} losses that favor fewer output sources and a \emph{covariance} loss that discourages correlated outputs.
We also experiment with a semantic \emph{classification} loss by predicting weak class labels for each mixture.
To handle larger numbers of sources, we introduce an efficient approximation using a fast least-squares solution, projected onto the MixIT constraint set. Our experiments show that the proposed losses curtail over-separation and improve overall performance. The best performance is achieved using larger numbers of output sources, enabled by our efficient MixIT loss, combined with sparsity losses to prevent over-separation. On the FUSS test set, we achieve over 13~dB in multi-source SI-SNR improvement, while boosting single-source reconstruction SI-SNR by over 17~dB.
\end{abstract}


\section{Introduction}
\label{sec:intro}
A basic problem in auditory perception is that sounds are superimposed, creating a notoriously difficult inverse problem: in a single-channel recording there are more unknown variables in the source samples than there are known variables in their mixture.
In recent years,  great strides have been made in source separation using supervised deep learning, 
raising the bar in performance on tasks such as separation of speech from non-speech %
\cite{huang2014deep, weninger2015speech},  separation of speech from speech %
\cite{hershey2016deep,isik2016single,yu2017permutation}, and separating arbitrary sounds from each other regardless of their class (universal sound separation) \cite{kavalerov2019universal, tzinis2020improving}.   

Supervised  separation uses isolated source waveforms as ground-truth output targets,  synthetically mixing them to create training inputs.  This is necessary because it is not feasible to record isolated sources in the same environment as their natural acoustic mixtures.  As a result, supervised training only works well when the synthetic mixtures match the domain in which the system is intended to perform.
For universal separation, with an open domain of sound types, it is difficult to match realistic source and mixture characteristics.  Databases of isolated sources do not exist for all 
classes that may be of interest, and realistic patterns of source activity and acoustic conditions
are unknown and difficult to infer.  However, copious  mixture data are available: arguably the vast majority of audio data in existence are unlabeled acoustic mixtures.   

Recently, \emph{mixture invariant training} (MixIT) \cite{wisdom2020mixit}, showed strong performance  without relying on isolated ground-truth sources, by training directly on mixtures of unknown source content.  
MixIT uses mixtures, instead of isolated sources, as references and a mixture of these mixtures as its input. 
The model outputs estimated sources such that they can be recombined to reconstruct the reference mixtures.  

One outstanding issue with MixIT is that it can lead to models that \emph{over-separate}, in the sense of producing a greater number of active source estimates than there are true underlying sources \cite{wisdom2020mixit,wisdom2020fuss}.  This is because the MixIT loss is invariant to the number of active estimated sources, as long as they can be remixed to approximate the reference mixtures with the same error.  When it is uncertain which components belong to the same mixture, a model may take advantage of MixIT's oracle assignment of sources to mixtures, and separate those components without penalty, avoiding the risk that they belong to different mixtures;  this provides an incentive for over-separation.  
  
Over-separation leads to lower separation fidelity with respect to the individual sources.  An unmistakable sign  is poor reconstruction when a single source is used as input: the source tends to be split across multiple outputs
\cite{wisdom2020mixit}.    
Including some supervised training data can reduce over-separation  \cite{wisdom2020mixit},
but %
the problem remains in the unsupervised setting, where isolated sources are unavailable.  

The over-separation issue worsens %
as the number of estimated sources increases. %
We introduce several new losses to address this.  A \emph{sparsity} loss encourages the pattern of activity across sources to be sparse.  A \emph{covariance} loss encourages the model to output decorrelated sources.  
 Finally, we experiment with a \emph{classification} loss that utilizes weak class labels for each mixture.  This encourages the semantic labels for each source to be different from each other, while matching the ensemble labels for the mixture.  

To carry out this investigation across a wider range of output sources $M$, we introduce an efficient version of MixIT that avoids brute force search \cite{wisdom2020mixit} %
and enables MixIT to be used with much larger numbers of output sources.
Our experiments show that both sparsity and classification losses 
curtail over-separation, and produce better multi-source separation performance. The best performance is achieved using 16 output sources combined with sparsity losses.

\section{Methods}
\label{sec:model}

We train separation networks using the same architecture as previous works \cite{kavalerov2019universal,wisdom2020mixit,wisdom2020fuss,wang2021sequential}, which separates sources by masking in a learned transform domain.
The network is composed of a learnable encoder/decoder with 2.5~ms window and 1.25~ms hop, combined with a time-domain convolutional network (TDCN++). The TDCN++ takes the encoder coefficients as input, and predicts $M$ masks through a sigmoid activation. The masks are elementwise multiplied with the encoder coefficients, and $M$ time-domain sources are reconstructed using the learnable decoder.
A mixture consistency projection \cite{wisdom2018consistency} ensures these sources sum to the input waveform.

\vspace{-2pt}
\subsection{Mixture invariant training (MixIT)}

In mixture-invariant training, inputs are formed by summing  $N$ reference mixtures $x_n\in\mathbb{R}^T$ to a mixture of mixtures (MoM),
$%
  \bar{x} = \sum\nolimits_n x_n.
$ %
The separation model $f_\theta$ predicts $M$ sources $\mathbf{\hat{s}}\in\mathbb{R}^{M\times T}$ from the MoM:
$%
  \hat{\mathbf{s}} = f_\theta(\bar{x})
$. %
Given reference mixtures and separated sources, the MixIT loss \cite{wisdom2020mixit} estimates a mixing matrix $\mathbf{A}\in\mathbb{B}^{N \times M}$:
\begin{align}
\mathcal{L}_\mathrm{MixIT}
\left(
    \{{x}_n\}, \hat{\bf s}
\right)
=
\min_{{\bf A}\in\mathbb{B}^{N \times M}} \, & \sum_n\mathcal{L}
    \left(
        {x}_n, [{\bf A} \hat{\bf s}]_n
    \right)
\label{eq:mixit}
\end{align}
where $\mathbb{B}^{N \times M}$ is the set of $N\times M$ binary matrices where each column sums to $1$, i.e., the set of matrices that assign each separated source $\hat{s}_m$ to one of the reference mixtures ${x}_n$, and $\mathcal{L}$ is a signal-level loss function between reference mixtures and their estimates.
We use the negative thresholded SNR as the signal-level loss function:
\begin{equation}
    \mathcal{L}({y}, \hat{y})
    =-10\log_{10}
    \frac{\|{y}\|^2}
    {\|{y}-\hat{y}\|^2 + \tau \|{y}\|^2}
    \label{eq:snr}
\end{equation}
where $\tau=10^{-\maxsnr / 10}$ acts as a soft threshold that clamps the loss at $\maxsnr$.
We find $\maxsnr=30$ dB to be a good value.

\vspace{-2pt}
\subsection{Efficient MixIT}
In the original formulation of MixIT \cite{wisdom2020mixit}, exhaustive $\mathcal{O}(N^M)$ search was used to find $\mathbf{A}$ in (\ref{eq:mixit}).
This is tractable when $N$ and $M$ are small, but for larger values, even for $N=2$ and $M>8$, it quickly becomes infeasible. To address this problem, we propose an efficient version of MixIT. In this efficient version, we simply use least-squares to find the optimal real-valued mixing matrix between reference mixtures $\mathbf{x}$ and estimated sources $\mathbf{\hat{s}}$, then project %
to the nearest valid binary mixing matrix:
\begin{equation}
\begin{aligned}
\mathbf{\hat{A}}
=
\mathcal{P}_{\mathbb{B}}
\left\{
    \mathrm{argmin}_{\mathbf{A} \in \mathbb{R}^{N\times M}}\:
    \|\mathbf{x} - \mathbf{A}\mathbf{\hat{s}}\|_2^2
\right\},
\end{aligned}
\label{eq:mixit_eff}
\end{equation}
where the projection operator $\mathcal{P}_\mathbb{B}$ sets the maximum value in each column to $1$, and the rest of the elements to $0$.
We have found that this surprisingly simple method yields essentially equivalent results to using exhaustive search, at a fraction of the cost.

\subsection{Source sparsity loss}
To directly address over-separation, we experiment with losses that encourage sparsity.
Specifically,  we consider sparsity in the distribution of activity among the estimated sources, measured in terms of the root-mean-squared (RMS) level of the time domain signals:
\begin{align}
    r_m = \op{rms}(\hat{s}_m) \defeq \sqrt{\frac{1}{T}\sum\nolimits_t |\hat{s}_{m,t}|^2 }
\end{align}

The $\ell_1$-norm has a long history as a sparsity inducing regularizer, used on  coefficients in linear regression problems in  \emph{LASSO}~\cite{tibshirani1996regression}.  It has also been used as a prior on sources in blind source separation \cite{lee1999blind}.   The $\ell_2$-norm on the other hand is well known to favor uniformity over sparsity.
Here we consider the $\ell_1$ norm over sources as $\|\mathbf{r}\|_1 = \sum_m r_m$, with a loss defined as:
\begin{align}
    C_{\ell_1}(\mathbf{r}) \defeq \frac{\frac{1}{M}\|\mathbf{r}\|_1 }{\op{rms}(\bar{x})},
    \label{eq:sparsity_l1}
\end{align}
where the  denominator and $\nicefrac{1}{M}$ factors normalize the scale across different examples and numbers of sources. 
Note that $C_{\ell_1}(\mathbf{r})$ encourages all sources to "shrink" towards zero. However the mixture consistency constraint
prevents shrinkage overall, while allowing changes in amplitude among sources.  

A potential drawback of the $\ell_1$ norm in $C_{\ell_1}(\mathbf{r})$ is that, because of the way sources combine, it assigns a higher loss when merging sources that are correlated than when merging independent sources.  This also happens with the $\ell_2$ norm,
and motivates $\ell_1$ / $\ell_2$ as a loss:
\begin{align}
    C_{\ell_1/\ell_2}(\mathbf{r}) &= \frac{\frac{1}{M}\|\mathbf{r}\|_1}{\|\mathbf{r}\|_2}.
    \label{eq:sparsity_l1l2ratio}
\end{align}
In this loss, the preference for merging independent sources cancels out, leaving only a preference for merging sources.   
A similar loss has been used 
\cite{hoyer2004non,yin2014ratio,krishnan2011blind} to promote scale-invariant sparsity.    

\subsection{Covariance loss}
One over-separation failure mode might be when the same source is evenly divided among multiple outputs.  In another failure case, an artifact may be added to one source and subtracted from another, so that they cancel out in the mixture, but disrupt the clean separation of the sources.  To combat such possibilities, we introduce a covariance-based loss that encourages separated sources to be uncorrelated.
We define such a loss based on the $\ell_1$ norm of off-diagonal covariances:
\begin{align}
  C_{\mathrm{cov}}(\hat{s}) &= \sum_{m\neq m'}|\op{cov}(\hat{s}_m, \hat{s}_{m'})|.
    \label{eq:cov}
\end{align}
\vspace{-15pt}

\begin{figure*}[th]
    \centering
    \includegraphics[width=0.49\linewidth, trim=0 0.9cm 0 0, clip ]{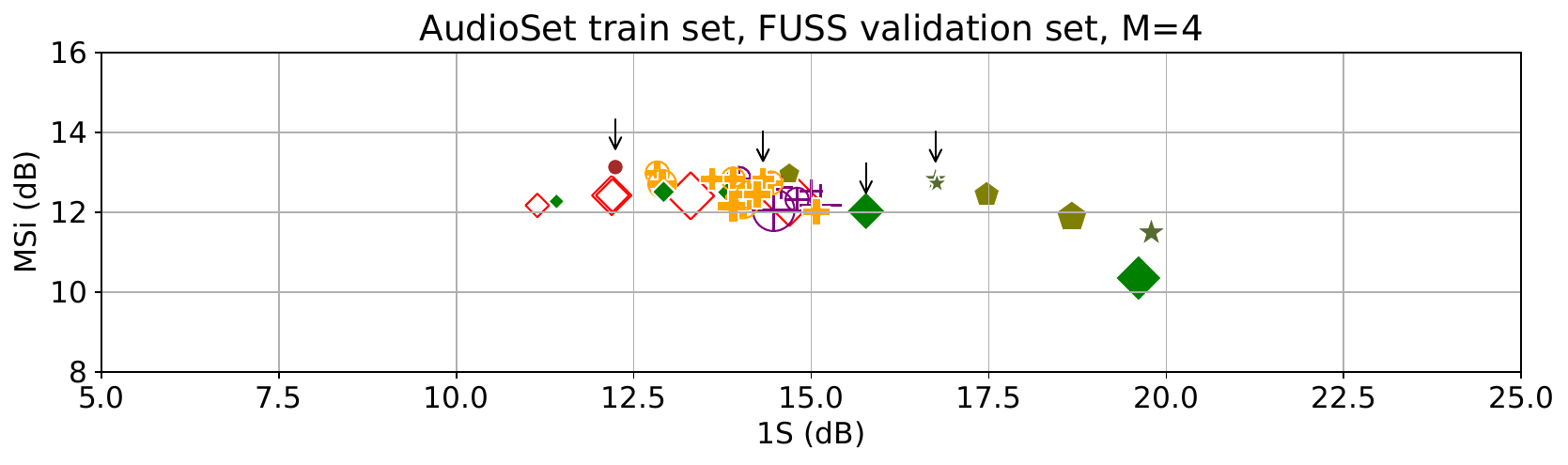}
    \includegraphics[width=0.49\linewidth, trim=0 0.9cm 0 0, clip]{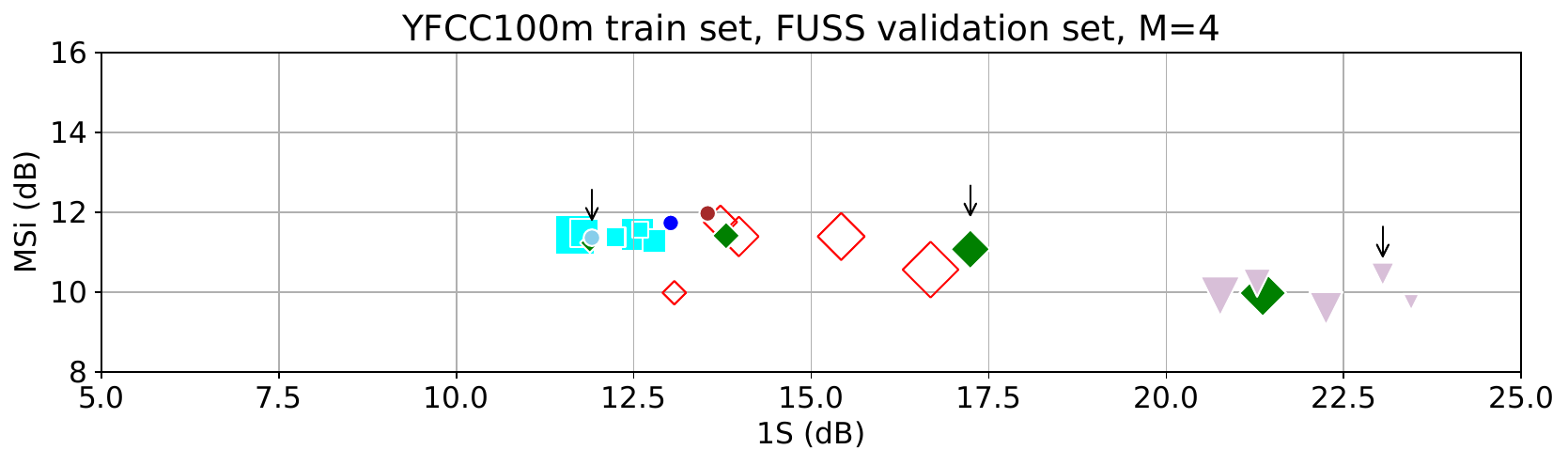}
    \\
    \includegraphics[width=0.49\linewidth]{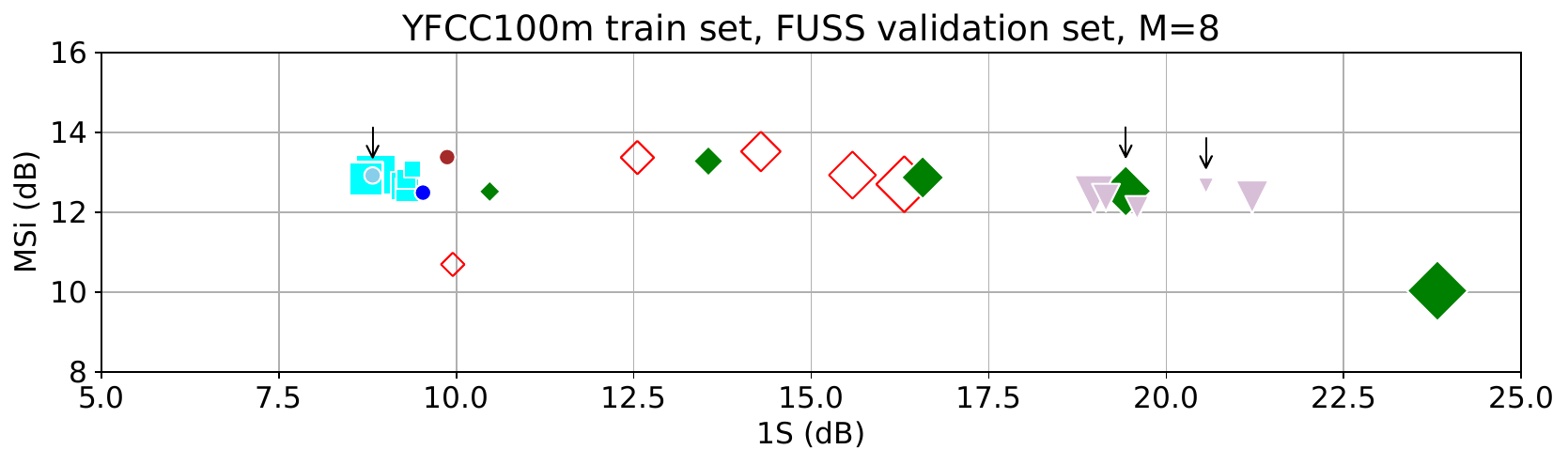}
    \includegraphics[width=0.49\linewidth]{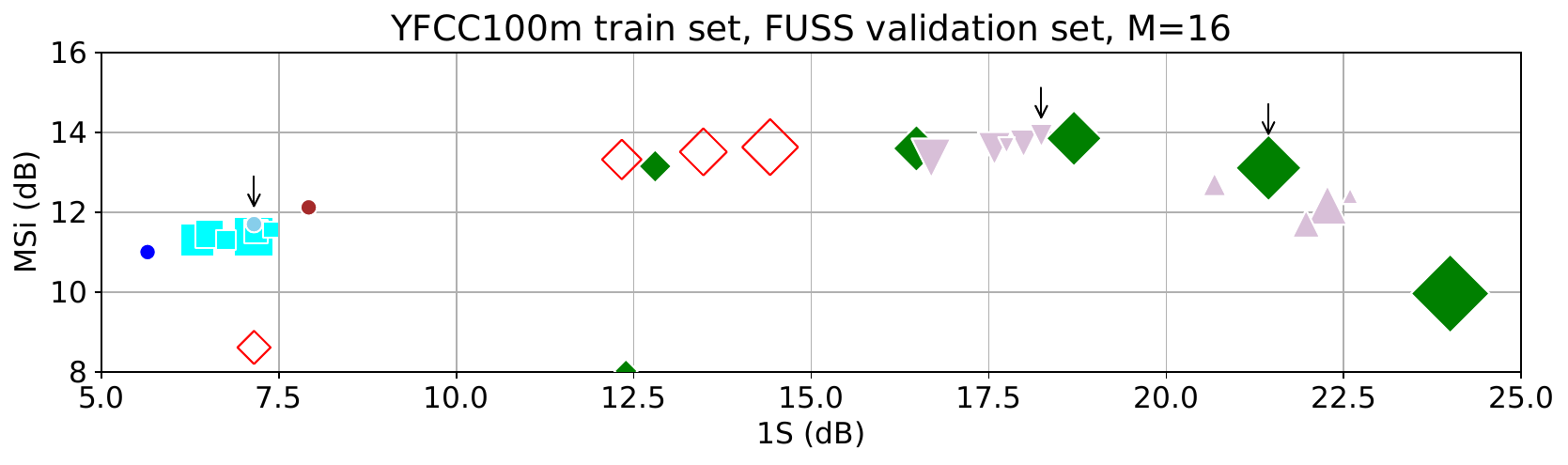}
    \\
    \centering
    \includegraphics[width=\linewidth]{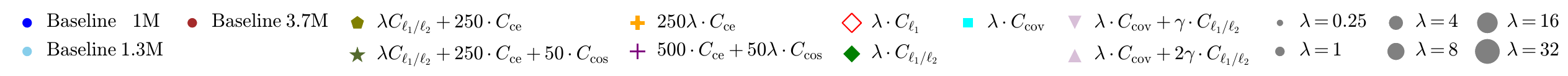}
    \\
    \vspace{-10pt}
    \caption{Scatter plots of MSi versus 
    1S for various losses. Marker size is proportional to the weight $\lambda$.
    By default all classification losses $C_\mathrm{ce}$ use OR aggregation, except those denoted with $\oplus$, which use XOR. 
    For $C_\mathrm{cov}$ experiment sweeps we used $\gamma=8\sqrt{M}$.
    Arrows denote hyperparameter settings which maximize 
    TRF, as depicted in Table~\ref{tbl:results}.
    }
    \label{fig:scatter}
    \vspace{-12.5pt}
\end{figure*}

\subsection{Joint separation and classification}
Even when clean source recordings are not available for supervised training, semantic labels may be available for the mixtures.  These labels provide two constraints on the separated sources.  First, assuming exhaustive labeling, the union of the classes that are present across the individual sources should match the semantic labels for the original mixture.  Second, when the sources are semantically distinct, the mutually exclusive labels for the original mixture should be distributed across separate sources.  These constraints can be used to combat over-separation  by feeding the separated outputs into a downstream classifier and using its predictions to define additional losses.

We begin by applying a pretrained $K$-class sound event classifier, $g: \mathbb{R}^T \rightarrow [0, 1]^K$, that maps each of the $M$ separated sources $\hat{s}_m$ to a corresponding posterior vector $p_m = g(\hat{s}_m)$.  Our first loss attempts to align the union of predictions with the ground truth labels provided for the mixture.  This is implemented through (i)~the aggregation of posterior vectors across separated sources into a single prediction for the mixture, and (ii)~the classification loss of that aggregate prediction and the ground truth labels.  This cross entropy loss takes the form (assuming multi-labeled data):
\begin{equation}
C_\mathrm{ce}(\mathbf{p}) = -\sum _k l_k \log \mathcal{A}(\mathbf{p})[k] + (1-l_k) \log(1-\mathcal{A}(\mathbf{p})[k]),
\end{equation}
where $\mathcal{A}$ is an aggregation function mapping $\mathbf{p}=\{p_m\}_{m=1}^M$ to a single posterior vector for the mixture, and $l_k \in \{0,1\}$ is the ground truth label for the $k$-th class.  We consider two aggregation functions.  The first is a soft logical-OR function, $\mathcal{A}_{\mathrm{or}}(\mathbf{p}) = 1 - \prod_{m=1}^M (1-p_m)$, which allows any source to constructively provide evidence for a particular class with no penalty for cross-source redundancy.  The second is a soft logical-XOR
function, $\mathcal{A}_\mathrm{xor}(\mathbf{p}) = \sum_{m=1}^M p_m \prod_{m' \neq m}^M (1-p_{m'})$, which encourages only one source to provide evidence for each class  present in the  mixture.

Both aggregation functions encourage semantically guided separation of the composite sound sources. $\mathcal{A}_\mathrm{xor}$ further encourages each semantic category to be restricted to a single separation output.  However, XOR aggregation does not explicitly require that each class is present on a \emph{different} separation output.  To encourage this behavior, we introduce a second average cosine similarity loss between pairs of output source posterior vectors, given by
\vspace{-2.5pt}
\begin{equation}
C_\mathrm{cos}(\mathbf{p}) = \frac{2}{M(M-1)}\sum_{m,m' > m}\Big(1+\frac{p_m^T p_{m'}}{\|p_m\| \|p_{m'}\|}\Big).
\label{eq:cos}
\end{equation}
\vspace{-2.5pt}
\noindent
This explicitly encourages posterior vectors to be orthogonal, which encourages both semantic mutual exclusivity of the separated outputs (like XOR) and distribution of semantic content across more sources.

\section{Experiments}
\label{sec:experiments}

\begin{table*}[th]
\centering
\caption{Results on various test sets for models trained with PIT on supervised FUSS, and with MixIT on unsupervised YFCC100m or AudioSet. The best loss type and weight are chosen based on maximizing TRF
on the FUSS validation set.
The best test metrics for each train set are bold, and the best FUSS metrics that maximize TRF 
are bold and italic.
}
\setlength{\tabcolsep}{3.5pt}
\scalebox{0.85}{
\begin{tabular}{@{}p{0.6in}llr @{\hskip 6\tabcolsep} rrr@{\hskip\tabcolsep}r @{\hskip 6\tabcolsep} rr @{\hskip 6\tabcolsep} r@{}r @{\hskip 6\tabcolsep} r@{}}
\toprule
& & & & \multicolumn{4}{@{}c@{\hskip 6\tabcolsep}}{FUSS}
& \multicolumn{2}{@{}c@{\hskip 6\tabcolsep}}{Libri2Mix}
& \multicolumn{2}{@{}c@{\hskip 6\tabcolsep}}{Enhancement}
& \multicolumn{1}{c}{YFCC100m} \\
\cmidrule(r{\dimexpr 6\tabcolsep}){5-8}
\cmidrule(r{\dimexpr 6\tabcolsep}){9-10}
\cmidrule(r{\dimexpr 6\tabcolsep}){11-12}
\cmidrule(l){13-13}
Train set & Method & Best loss 
& $M$
& MSi & 1S & TRF & MoMi
& MSi & 1S
& S+N & S-only
& MoMi \\
\midrule
\multirow{1}{3pt}{FUSS (supervised)}
& Single-mixtures &  & 4 
&12.4 & {\bf 40.9} & 19.5 & 4.3 & -1.9 & 15.6 & -1.4 & 11.5 & -0.1 \\
&  (1-4 sources) & & 8 
&12.6 & 36.8 & 18.7  & 3.5 & -2.8 & {\bf 17.9} & -3.8 & 9.3 & -0.2 \\
&  &  & 16 &\textbf{\textit{13.7}} & \textbf{\textit{38.5}} & \textbf{\textit{19.9}}  & 4.7 & -2.6 & 17.8 & -1.3 & 12.3 & -1.9 \\
\addlinespace
& MoMs &  & 8 
&13.4 & 14.8 & 13.8  & 8.4 & {\bf -0.6} & 15.8 & 0.4 & {\bf 13.3} & {\bf 1.1} \\
& (2-8 sources) &  & 16 &{\bf 13.7} & 11.3 & 13.1  & {\bf 8.9} & -0.7 & 13.2 & {\bf 0.8} & 13.1 & 1.0 \\
\midrule
\multirow{1}{3pt}{YFCC100m (unsup.)}
& Baseline 1.3M &  & 4 
&11.9 & 13.8 & 12.4  & 7.4 & {\bf 10.0} & 24.5 & {\bf 9.3} & 28.6 & 8.5 \\
&  &  & 8 
&12.7 & 9.9 & 12.0  & 9.9 & 8.7 & 22.9 & 9.2 & 23.1 & 9.1 \\
&  &  & 16 
&11.0 & 8.2 &  10.3 & {\bf 10.4} & 5.8 & 20.6 & 9.2 & 23.3 & {\bf 9.4} \\
\addlinespace
& Sparsity & $8 \cdot C_{\ell_1 / \ell_2}$  & 4 
&11.6 & 20.7 & 13.9  & 6.3 & 9.3 & 31.4 & 8.7 & 30.6 & 8.0 \\
& 1M+0.3M & $23 \cdot C_{\ell_1 / \ell_2}$  & 8 
&12.4 & 23.9 & 15.3  & 6.4 & 9.7 & 34.2 & 8.8 & 31.0 & 8.1 \\
&  & $64 \cdot C_{\ell_1 / \ell_2}$  & 16 
&\textbf{\textit{13.2}} & \textbf{\textit{25.5}} & \textbf{\textit{16.3}}  & 6.0 & 8.9 & 34.9 & 7.4 & 32.0 & 7.4 \\
\addlinespace
& Sparsity + Cov. & $16 \cdot C_{\ell_1 / \ell_2} + 4 \cdot C_\mathrm{cov}$  & 4 
&10.2 & {\bf 26.3} & 14.2  & 4.4 & 9.3 & {\bf 38.0} & 7.9 & {\bf 35.8} & 7.6 \\
& 1M+0.3M & $23 \cdot C_{\ell_1 / \ell_2} + 1 \cdot C_\mathrm{cov}$  & 8 
&12.5 & 24.6 & 15.5  & 5.6 & 9.3 & 34.7 & 8.0 & 31.7 & 8.0 \\
&  & $64 \cdot C_{\ell_1 / \ell_2} + 0.25 \cdot C_\mathrm{cov}$  & 16 
&13.0 & 25.8 & 16.2  & 5.4 & 8.7 & 34.3 & 6.9 & 32.2 & 7.5 \\
\midrule
\multirow{1}{3pt}{AudioSet (unsup.)}
& Baseline 3.7M &  & 4 &{\bf 13.4} & 13.3 & 13.4  & {\bf 9.2} & 13.4 & 28.1 & 10.9 & 29.1 & {\bf 8.3} \\
\addlinespace
& Sparsity & $16 \cdot C_{\ell_1 / \ell_2}$  & 4 
&12.5 & 18.6 & 14.0  & 7.2 & 12.8 & {\bf 29.5} & 10.6 & 29.5 & 7.4 \\
& Class. & $250 \cdot C_\mathrm{ce}$  & 4 
&13.2 & 16.3 & 14.0  & 8.9 & 13.3 & 26.3 & 10.9 & 28.4 & 8.0 \\
& Sparsity + Class. & $16 \cdot C_{\ell_1 / \ell_2} + 250 \cdot C_\mathrm{ce}+50 \cdot C_\mathrm{cos}$  & 4 
&\textbf{\textit{13.1}} & \textbf{\textit{19.6}} & \textbf{\textit{14.7}}  & 8.3 & {\bf 13.6} & 29.2 & {\bf 11.1} & {\bf 30.1} & 7.6 \\
\bottomrule
\end{tabular}
}
\label{tbl:results}
\vspace{-10pt}
\end{table*}

We train unsupervised separation models using MixIT on two large datasets of 10-second clips extracted from in-the-wild audio containing a wide variety of different source classes:
YFCC100m~\cite{thomee2016yfcc100m}, containing about 1600 hours of audio, and AudioSet~\cite{AudioSet},  containing about 5800 hours. Each clip is annotated with semantic class labels describing the sound events contained therein. All models are fine-tuned from a baseline checkpoint trained only with the MixIT loss.  The  Adam optimizer \cite{kingma2014adam} was used with batch size 256 and learning rate $10^{-3}$ on 4 Google Cloud TPUs (16 chips).
Exhaustive MixIT is used for $M \leq 8$, and efficient MixIT otherwise\footnote{For $M$ of 4 and 8, we ran experiments with efficient MixIT, and did not observe any significant difference compared to using the exhaustive MixIT.}.
Average wall-clock time for 300k steps was M=4: 1.5 days, M=8: 2.5 days, and M=16: 3 days.
For YFCC100m, we initialize from a checkpoint trained to 1M steps and fine-tune for an additional 300k steps using various losses. For AudioSet, we use a checkpoint trained to 3.7M steps and fine-tune %
using early stopping.
To apply classification losses, we feed the separation outputs into a pretrained ResNet-18 AudioSet classifier that processes waveform inputs with a simple learned front end~\cite{sainath2015}. We found it necessary to jointly train the classifier when using the cosine-similarity loss (\ref{eq:cos}), but that it could be held fixed when using soft-OR and soft-XOR losses alone. To fit separation and classification networks into memory, we used 3-second clips.

Separation performance is evaluated using several supervised synthetic datasets containing reference sources.
Since we train our separation models on a universal sound separation task, we primarily focus on evaluation with the FUSS dataset~\cite{wisdom2020fuss}, which contains 10-second mixtures of one to four arbitrary sound sources drawn from 
\href{https://www.freesound.org}{\texttt{freesound.org}}. 
Since many practical separation applications focus on speech signals, we additionally evaluate how well unsupervised universal separation models can generalize to two  specific task domains:
speech separation, using mixtures of two overlapping speakers from Libr2Mix~\cite{cosentino2020librimix},
and speech enhancement, using the same dataset as previous work  \cite{wisdom2020mixit} in which speech is drawn from 
\texttt{librivox.org},
and nonspeech from 
\texttt{freesound.org}.

We use several separation metrics, as proposed for measuring performance on FUSS \cite{wisdom2020fuss} and for MixIT-trained models \cite{wisdom2020mixit}:
\begin{itemize}[leftmargin=*]
\item {\bf MSi}: for %
mixture inputs with 2+ sources, scale-invariant SNR \cite{LeRoux2018a} improvement (SI-SNRi) in dB relative to the input for all active reference-estimate pairs.
\item {\bf 1S}: the reconstruction SI-SNR in dB, using a single output source, when passing a single source through the separation model (SI-SNRi  is not useful here because the input has infinite SI-SNR).
\item {\bf TRF}: total reconstruction fidelity, a weighted sum of 1S and  MSi   ($\operatorname{TRF} = p_1 \mathrm{1S} + \sum_{m=2}^M p_m \mathrm{MSi}_m$, where $p_m$ is the proportion of data with $m$ sources, and $\mathrm{MSi}_m$ is the MSi for $m$-source data.)
\item {\bf MoMi}: the SI-SNRi in dB of reconstructed \emph{mixtures} found using the optimal binary mixing matrix ${\bf A}$ using to align the ground truth to the separated sources.
This is the same as the loss function in \eqref{eq:mixit}, and is the only objective measure we can compute on unsupervised datasets which do not contain isolated sources.
\end{itemize}
For speech enhancement, we only measure SI-SNR improvement of the target speech source, ignoring the nonspeech source. In Table~\ref{tbl:results}, speech SI-SNRi for noisy speech mixtures is referred to as {\bf S+N}, and for clean speech input as {\bf S-only}.
All metrics are computed with permutation invariance using the $\mathcal{O}(M^3)$ Hungarian algorithm~\cite{kuhn1955hungarian}, since brute-force $\mathcal{O}(M!)$ alignment is infeasible for higher $M$.

Figure \ref{fig:scatter} visualizes the trade-off between MSi versus 1S on the FUSS validation set for various $M$ over the set of regularization losses, with weight $\lambda$ swept over a wide range.  
Marker sizes are proportional to $\sqrt{\lambda}$. Notice that for the sparsity losses, as $\lambda$ increases, MSi and 1S generally continue to increase up to a critical point.
Past this critical point the sparsity losses become too strong, causing the models to under-separate: 1S continues to increase while MSi decreases.
Across tasks, we find that using the $C_{\ell_1/\ell_2}$ sparsity loss \eqref{eq:sparsity_l1l2ratio} consistently improves 1S compared to $C_{\ell_1}$ \eqref{eq:sparsity_l1}.
Covariance loss \eqref{eq:cov} alone does not improve over the baseline; however, combining covariance loss with sparsity leads to small improvements in 1S.

A summary of the best results on the test sets for different $M$ is provided in Table~\ref{tbl:results}, along with supervised baselines using the permutation invariant training (PIT) procedure used in the original FUSS separation model \cite{wisdom2020fuss} on either single mixtures (1-4 sources) or MoMs (2-8 sources) from FUSS. Note that the optimal $\lambda$ increases with $M$ because over-separation in the baseline tends to increase with $M$.
For each loss type, the checkpoint and loss weight that maximizes TRF
on the FUSS validation set is selected,  
which
reflects the proportion of multi-source to single-source mixtures.

Our best results on FUSS, in terms of maximizing TRF, are achieved by a $M\!=\!16$ model trained on YFCC100m using the sparsity loss \eqref{eq:sparsity_l1l2ratio}.
This model achieves a new state-of-the-art for unsupervised models on the FUSS test set: 13.2 dB MSi and 25.5 dB 1S, which is only 0.5 dB MSi behind the fully supervised baseline, and boosts 1S by over 17 dB compared to the unsupervised $M=16$ baseline's 8.2 dB 1S.
Incorporating covariance loss \eqref{eq:cov} improves over sparsity alone in terms of 1S, with a slight reduction in MSi. 

In general, we find that all unsupervised MixIT models consistently avoid over-separating speech sources, as evidenced by high 1S scores on Libri2Mix and enhancement; however, speech MSi is improved with additional regularization. Supervised FUSS models perform very poorly on speech separation MSi and enhancement S+N, likely due to the limited amount of  source data and diversity in FUSS.
This highlights an advantage of training on a large amount of real-world data, spanning a variety of source types.
Finally, note that incorporating sparsity losses consistently hurts MoMi on all tasks. We hypothesize this is caused by more sources being forced to zero, which makes it more difficult to reconstruct reference mixtures.

As shown in Table~\ref{tbl:results}, the AudioSet baseline trained for 3.7M steps is quite strong, achieving better scores on FUSS compared to the YFCC100m $M=4$  model trained for 1.3M steps.
Sparsity regularization alone improves 1S at a cost of almost 1 dB MSi.
The best classification loss setting improves 1S significantly, with minimal degradation in MSi and scores well on other tasks, indicating its ability to reduce over-separation.  %
Our best results from AudioSet training are produced by combining the classifier soft-OR and cosine similarity losses and sparsity regularization (bottom row).

Finally, note the improved performance on speech tasks using AudioSet models, despite being trained on unmatched data without source-level supervision. In particular, baseline MSi on Libri2Mix speech separation (13.4 dB) comes surprisingly close to a fully supervised model trained only on matched data (16 dB) \cite{cosentino2020librimix}.
This reflects the fact that about half of AudioSet clips contain speech \cite{AudioSet}.

Please see the project website \cite{waspaa2021_audio_demos} for audio demos.

\section{Conclusion}

We have presented a number of auxiliary losses that improve unsupervised sound separation models trained using MixIT, especially for models which emit many output sources. These losses include explicit sparsity penalties on the magnitudes of separated sources and their covariances, as well as semantic classification losses.  To enable training with a large number of output sources, we proposed an efficient least-squares-based MixIT implementation.
Used in combination with sparsity regularization, \emph{unsupervised} models achieved state-of-the-art performance on universal sound separation, in terms of MSi and 1S on the FUSS test set. These models are also suitable for general purpose use, demonstrating good performance on domain-specific tasks such as speech separation and enhancement. In future work, we plan to scale up model and data size to train even better general-purpose  separation systems, as well as continue to explore combinations of sound separation and classification systems.

\ifincludeacknowledgements
\section{Acknowledgments}
Thanks to Kevin Wilson for helpful comments, and to Jiquan Ngiam for TensorFlow implementation of the Hungarian algorithm.
\fi

\bibliographystyle{IEEEtran}
\bibliography{refs}

\end{sloppy}
\end{document}